\documentclass{aa}

\usepackage{graphicx}
\usepackage{txfonts}
\usepackage{orcidlink}
\def\epjc{EPJC}

\begin{document}

   \title{Vacuum breakdown around a Kerr black hole surrounded by a magnetic field}

   \author{C. Cherubini \inst{1,2} \orcidlink{0000-0002-0542-5601}
        \and R. Moradi \inst{3} \orcidlink{0000-0002-2516-5894}  
        \and J. A. Rueda \inst{2,4,5,6,7} \orcidlink{0000-0003-4904-0014} 
        \and R.~Ruffini \inst{2,4,8} }

\institute{Department of Science and Technology for Sustainable Development and One Health, Università Campus Bio-Medico di Roma, Via Alvaro del Portillo 21, I-00128 Rome, Italy \\
\email{c.cherubini@unicampus.it}
\and
ICRANet, Piazza della Repubblica 1, I-65122 Pescara, Italy\\
\email{ruffini@icra.it}
\and
Key Laboratory of Particle Astrophysics, Institute of High Energy Physics, Chinese Academy of Sciences, 100049 Beijing, China\\
\email{rmoradi@ihep.ac.cn}
\and
ICRA, Dipartimento di Fisica, Sapienza Università di Roma, I-00185 Rome, Italy
\and
ICRANet-Ferrara, Dip. di Fisica e Scienze della Terra, Università degli Studi di Ferrara, Via Saragat 1, I-44122 Ferrara, Italy\\
\email{jorge.rueda@icra.it}
\and
Dipartimento di Fisica e Scienze della Terra, Università degli Studi di Ferrara, Via Saragat 1, I-44122 Ferrara, Italy\\
\and
INAF, Istituto de Astrofisica e Planetologia Spaziali, Via Fosso del Cavaliere 100, I-00136 Rome, Italy\\
\and
INAF, Viale del Parco Mellini 84, I-00136 Rome, Italy
}

\date{Received \today / Accepted }

% ---------------  ABSTRACT ------------   

\abstract{
We present the invariant characterization of the region where vacuum breakdown into electron-positron ($e^+e^-$) pairs occurs due to an overcritical electric field, the \textit{dyadoregion}, in the case of a Kerr black hole (BH) in the presence of an external, asymptotically uniform test magnetic field aligned with the BH rotation axis, using the Wald solution. We calculate the dyadoregion morphology, the electromagnetic energy available for the pairs, the pair-creation rate, the number density of pairs, the average energy per pair, and their energy density and pressure. These results provide initial conditions for simulating the subsequent dynamics of the pair-produced plasma and astrophysical applications in the context of high-energy transients involving BHs in strong electromagnetic fields.
}

 %-- 

\titlerunning{Vacuum breakdown around a Kerr black hole surrounded by a magnetic field}
\authorrunning{Cherubini, et al.}

\maketitle

%%%%%%%%%%%%%%%%%%%%%%%%%%%%%%%%%%%%%%%%%%%%%%%%%%%%%%%%%%%
%%%%%%%%%%%%%%%%%%%%%%%%%%%%%%%%%%%%%%%%%%%%%%%%%%%%%%%%%%%
\section{Introduction} \label{sec:1}
%%%%%%%%%%%%%%%%%%%%%%%%%%%%%%%%%%%%%%%%%%%%%%%%%%%%%%%%%%%
%%%%%%%%%%%%%%%%%%%%%%%%%%%%%%%%%%%%%%%%%%%%%%%%%%%%%%%%%%%

The ultrarelativistic expansion and transparency of an electron-positron-photon ($e^+e^-\gamma$) plasma in a poorly baryon-contaminated medium has been considered since the early gamma-ray burst (GRB) discovery as a key ingredient to explain their prompt emission \citep[see, e.g.,][and references therein]{1975PhRvL..35..463D,2004RvMP...76.1143P,2015PhR...561....1K,2018pgrb.book.....Z}. The thermalization of the pair plasma is achieved on a short timescale of $\sim 10^{-13}$~s \citep{2007PhRvL..99l5003A,2009PhRvD..79d3008A}. The integration of the equations of motion of the initially optically thick plasma reveals that it self-accelerates and reaches transparency at ultra-relativistic speeds \citep{1999A&A...350..334R, 2000A&A...359..855R}. The dynamics depends on the baryon load parameter, $B \equiv M_B c^2/E_{e^+e^-}$, the ratio of the baryon's rest mass to the plasma energy. For low baryon load systems, $B \lesssim 10^{-2}$, the Lorentz factor at transparency is $\Gamma \sim 1/B$, so the plasma reaches $\Gamma \gtrsim 100$ \citep{1999A&A...350..334R,2000A&A...359..855R}.

The plasma dynamics depend on the initial and ambient conditions, which are crucially related to the origin of the $e^+e^-$ pair plasma. Traditional GRB models often use neutrino-dominated accretion flows \citep[called NDAFs][]{1999ApJ...518..356P} in which pairs formed around the BH due to the annihilation of thermally-produced neutrinos and antineutrinos emitted by the massive accretion disk of a collapsar \citep[see, e.g.,][]{1999ApJ...518..356P,2001ApJ...557..949N,2002ApJ...577..311K,2002ApJ...579..706D,2005ApJ...629..341K,2005ApJ...632..421L,2006ApJ...643L..87G,2007ApJ...657..383C,2007ApJ...662.1156K,2010A&A...509A..55J,2013ApJ...766...31K,2013MNRAS.431.2362L,2013ApJS..207...23X,2017NewAR..79....1L}. The low energy deposition of $\nu\bar{\nu}$ annihilation into $e^+e^-$ pairs leads these models to advocate for very high accretion rates (e.g., $\gtrsim M_\odot$ s$^{-1}$) to explain energetic GRBs \citep[see, e.g.,][and references therein]{2007ApJ...657..383C,2017NewAR..79....1L}. The energy deposition rate is even lower if neutrino flavor oscillations are considered \citep{2021Univ....7....7U}.

We here focus on the different possibility of $e^+e^-$ pair plasma formation around a rotating compact object due to nonlinear quantum electrodynamics (QED) vacuum breakdown by electric fields exceeding the critical value $E_c = m_e^2 c^3/(e \hbar) \approx 1.32\times 10^{16}$ V cm$^{-1}$. We refer to \citep{2010PhR...487....1R} for a topical review. 

The relevance of QED pair creation in GRBs has been clear from the first estimate presented for the pair creation around a charged, rotating BH by T. Damour and R. Ruffini in 1975 \citep{1975PhRvL..35..463D} using the Kerr-Newman spacetime \citep{1965JMP.....6..918N}. The systematic application of the QED vacuum polarization to GRB analysis started with the introduction of the concept of \textit{dyadosphere} \citep{1998A&A...338L..87P} of a charged BH using the Reissner-Nordstr\"om spacetime \citep{1916AnP...355..106R,1918KNAB...20.1238N}. The dyadosphere is the spherical region above the BH horizon where the electric field is larger than $E_c$. Thus, the dyadosphere surface radius is given by the electric field contour $E(r_d) = E_c$ (see Section \ref{sec:2} below), i.e., $r_d = \sqrt{Q/E_c}$, where $Q$ is the BH charge. In \citet{1998A&A...338L..87P}, the pair density, the number of pairs, average energy, and the plasma temperature were estimated. The thermodynamical properties of the pair plasma are the initial conditions that feed the equations of motion that drive the plasma dynamics, and whose transparency is relevant for the GRB prompt emission explanation \citep{1999A&A...350..334R,2000A&A...359..855R,2021PhRvD.104f3043M,2022EPJC...82..778R}. 

The dyadosphere concept has been extended to \textit{dyadoregion}, which considers the possibility of more general geometries than the spherical. For instance, it has been shown that the dyadoregion of the Kerr-Newman BH can assume a torus-like shape, a \textit{dyadotorus} \citep[see, e.g., ][for details]{2009PhRvD..79l4002C}.

The non-zero charge of the BH has been essential for the development of the dyadosphere or the dyadotorus. For instance, the existence of a dyadosphere implies the BH charge to be $Q > r_+^2 E_c$, hence a charge-to-mass ratio $Q/M \gtrsim 4 M E_c \approx 7.5\times 10^{-6} (M/M_\odot)$. However, astrophysical BHs are commonly adopted as neutral objects, assuming that charged particles of opposite charge would rapidly screen the BH charge as it captures them from the environment. This view is modified by two complementary aspects that can be recognized from the Wald solution \citep{1974PhRvD..10.1680W} of the Einstein-Maxwell equations of a Kerr BH (of mass $M$ and angular momentum $J$; \citealp{1963PhRvL..11..237K}), in the presence of an external, asymptotically uniform and aligned (with the BH angular momentum) test magnetic field of strength $B_0$ (see Appendix \ref{app:A}). First, the interaction of the BH rotation and the magnetic field induces a quadrupolar electric field, while the BH remains uncharged \citep[see, e.g.,][]{2000NCimB.115..751M,2022ApJ...929...56R}. Second, the induced electric field can attract or repel charged particles (depending on their charge sign), making a Kerr BH in a magnetic field an excellent site for producing radiation by charged particle acceleration, which is relevant, for instance, in the high-energy (GeV) emission of GRBs \citep{2019ApJ...886...82R,2020EPJC...80..300R,2021A&A...649A..75M,2022ApJ...929...56R}. 

From the above discussion arises a clear corollary: a rotating BH does not need to be charged to produce a dyadoregion and create a pair-photon plasma. This is the situation of interest in this article. A Kerr BH in such an environment could arise from the collapse of a magnetized neutron star (NS). The magnetic field around the BH can be inherited from the collapsed NS magnetic field, which could be amplified in the process \citep{2013PhRvD..88d4020D,2017MNRAS.469L..31N,2018ApJ...864..117M,2020ApJ...893..148R}, even to overcritical values. Still, it can be considered a test magnetic field as its strength fulfills $B_0 M \ll 1$, namely, $B_0 \ll 2.4\times 10^{19} (M_\odot/M)$ G.

Stellar-mass Kerr BHs in the presence of strong magnetic fields have been considered in the description of the GRB prompt emission, e.g., of GRB 190114C \citep{2021PhRvD.104f3043M} and GRB 180720B \citep{2022EPJC...82..778R}. There, the pair plasma parameters were estimated using the Kerr-Newman BH solution with an effective charge, $|Q_{\rm eff}| = 2 J B_0$. The reason behind using this effective charge is that, in the Wald solution, the induced electric field on the polar axis behaves as a field produced by a charge monopole of this value \citep{2019ApJ...886...82R,2021PhRvD.104f3043M} (see also Appendix \ref{app:A}). The dyadoregion energy was estimated using the formula for the electromagnetic energy stored in a spherical region around the Kerr-Newman BH, as provided in \citet{2009PhRvD..79l4002C}. The radius was set to $r_d = \sqrt{|Q_{\rm eff}|/E_c} = \sqrt{2 J B_0/E_c}$, which gives, for instance, $r_d = 4.75 M$ for a Kerr BH with $J = 0.5 M^2$ and an external magnetic field $B_0 = 10^{15}$ G. For these values, the effective charge is $|Q_{\rm eff}|/M = 8.46 \times 10^{-6} (M/M_\odot)$. Notice that, quantitatively, the effective charge is similar to the charge of the Reissner-Nordstr\"om BH considered in the dyadosphere analyses. The reason for this result is that, in both cases, the electric field is requested to be overcritical. Additional work on the possibility of pair creation in the context of the Wald solution can be found in \citet{2000PhRvL..84.3752V,2001PhRvD..63f4028H}.

Investigating whether the BH, with its surrounding material, can sustain strong magnetic fields remains an active topic of research \citep[see, e.g.,][]{2021PhRvL.127e5101B}, but falls beyond the scope of this work. In this line, it is worth mentioning that a co-rotating strong magnetic field could also induce overcritical electric fields around a fast-rotating, magnetized NS \citep[see, e.g., ][for recent three-dimensional simulations]{2024ApJ...976...80B}. Because the multipole moments of a rotating NS approaching the critical mass get close to the ones of a Kerr BH \citep{2015PhRvD..92b3007C}, likewise the exterior spacetime properties \citep{2017PhRvD..96b4046C}, we expect with some confidence that our main conclusions will remain valid, within some cautious level of approximation, also in that situation.

While the density and magnetic field in the surroundings of the rotating compact object can be very complex \citep[see, e.g.][]{2024ApJ...976...80B}, and the solution of the problem may require advanced numerical general relativistic magnetohydrodynamic simulations, our aim here is to provide an accurate characterization of the dyadoregion within a relatively simplified model, which highlights the main features and parameters relevant for the physical process. For this task, we confine ourselves to an analytical characterization of the dyadoregion in the Wald solution, without approximations. We shall also derive the thermodynamic properties of the pair-photon plasma, which serve as initial conditions for numerical simulations of the subsequent plasma dynamics. 

The article is organized as follows. In section \ref{sec:2}, we define the dyadoregion in a coordinate-independent way using the electromagnetic invariants and determine its morphology and energetics. Section \ref{sec:3} uses the above information to determine the thermodynamic properties of the associated pair plasma. Finally, section \ref{sec:4} summarizes the results of this article and discusses some astrophysical consequences. Mathematical and technical details are presented in Appendix \ref{app:A}, \ref{app:B}, and \ref{app:C}

%%%%%%%%%%%%%%%%%%%%%%%%%%%%%%%%%%%%%%%%%%%%%%%%%%%%%%%%%%%%
%%%%%%%%%%%%%%%%%%%%%%%%%%%%%%%%%%%%%%%%%%%%%%%%%%%%%%%%%%%%
\section{Defining and characterizing the dyadoregion} \label{sec:2}
%%%%%%%%%%%%%%%%%%%%%%%%%%%%%%%%%%%%%%%%%%%%%%%%%%%%%%%%%%%%
%%%%%%%%%%%%%%%%%%%%%%%%%%%%%%%%%%%%%%%%%%%%%%%%%%%%%%%%%%%%

One of the tasks of astrophysical interest is estimating the electromagnetic energy available for the pair creation. Given a $t=$ constant hypersurface, the electromagnetic energy stored inside the dyadoregion is [see Eq. (\ref{eq:energyfinal}) in Appendix \ref{app:B}]
\begin{equation}\label{eq:energyfinal2}
    {\cal E} = 2 \pi\int_0^\pi\int_{r_+}^{r(\theta)} \left(U_{\rm em} + S_{\hat 3} \frac{A \sin\theta}{\Sigma\sqrt{\Delta}}\right)  \Sigma\sin\theta dr d\theta,
\end{equation}
where $U_{\rm em} = (\hat{E}^2 + \hat{B}^2)/(8\pi)$ is the electromagnetic energy density, with $\hat{E}^2 \equiv E_{\hat{i}}E^{\hat{i}}$ and $\hat{B}^2 \equiv B_{\hat{i}}B^{\hat{i}}$ [see Eq. (\ref{eq:EBnorms})], and $S_{\hat 3} = |{\bf \hat E}\times {\bf \hat B}|/(4\pi) = (E_{\hat{1}} B_{\hat{2}} - E_{\hat{2}} B_{\hat{1}})/(4 \pi)$ is the Poynting vector, being $E_{\hat{i}}$ and $B_{\hat{i}}$ the electric and magnetic field components measured in the locally non-rotating frame (LNRF; see Appendix \ref{app:A}, for details). The integral is carried out over the surface $r(\theta)$, in our case of interest, the \textit{dyadoregion}. 

It is clear that, to perform the integral (\ref{eq:energyfinal2}), we need a mathematical definition of the dyadoregion. In words, one can define the latter as the region around the BH where vacuum polarization occurs. We now translate the above definition into a mathematical equation of the dyadoregion surface.

In his seminal paper, \citet{1951PhRv...82..664S} derived the rate of pair creation per unit four-volume in terms of the electromagnetic invariants 
\begin{equation}\label{eq:invariants}
{\cal F}\equiv\frac14F_{\mu\nu}F^{\mu\nu}=\frac12({\bf B}^2-{\bf E}^2),\quad 
{\cal G}\equiv\frac14F_{\mu\nu}{}^*F^{\mu\nu}={\bf E}\cdot {\bf B},
\end{equation}
where ${\bf E}$ and ${\bf B}$ are the electric and magnetic fields, clearly calculable in any frame as they are here employed to obtain the invariants. Assuming the fields are spatially uniform, Schwinger derived the {invariant} pair-creation number density rate
\begin{align}\label{eq:rate}
	{\frac{dN}{\sqrt{-g} d^4x}} &= \frac{\alpha\, {\mathcal G}}{4 \pi^2 \hbar}\sum_{{l}=1}^\infty\frac1{{l}}\coth\left[{l}\,\pi\sqrt{\frac{({\mathcal F}^2+{\mathcal G}^2)^{1/2}+{\mathcal F}}{({\mathcal F}^2+{\mathcal G}^2)^{1/2}-{\mathcal F}}}\right] \nonumber \\
    &\times \exp{\left[{-\frac{{l}\,\pi E_c}{\sqrt{({\mathcal F}^2+{\mathcal G}^2)^{1/2}-{\mathcal F}}}}\right]}\, ,
\end{align}
where {$g$ is the metric determinant, and} $\alpha$ the fine structure constant.  

Since the electric and magnetic fields surrounding the BH vary on macroscopic scales (e.g., kilometers for a stellar-mass BH), we can consider them uniform on the spatial scales where pair creation occurs, the electron Compton wavelength. This feature allows us to determine the local pair creation rate using Schwinger's treatment for uniform fields. 

Based on the electromagnetic invariants, the rate (\ref{eq:rate}) does not depend on the frame chosen to calculate the electric and magnetic fields\footnote{We refer to Appendix \ref{app:B} for a straightforward calculation of the electromagnetic invariants ${\cal F}$ and ${\cal G}$ via the Newman-Penrose formalism.}. However, it is interesting that this pair-creation rate acquires a simplified and appealing expression when using the electric and magnetic fields in a frame where they are parallel to each other. Let us denote the fields in that frame as ${\bf \tilde E}$ and ${\bf \tilde B}$. Thus, ${\cal G} = {\bf \tilde E} \cdot {\bf \tilde B} = \tilde E \tilde B$, where $\tilde E$ and $\tilde B$ are the field moduli. We can invert the system (\ref{eq:invariants}) to write the field moduli in terms of the electromagnetic invariants:
\begin{equation}\label{eq:PAREB2}
	\tilde E=[({\mathcal F}^2+{\mathcal G}^2)^{1/2}-{\mathcal F}]^{1/2},\quad \tilde B=[({\mathcal F}^2+{\mathcal G}^2)^{1/2}+{\mathcal F}]^{1/2}.
\end{equation}
Using Eq. (\ref{eq:PAREB2}), the pair production rate (\ref{eq:rate}) can be rewritten as
\begin{equation}\label{eq:rate2}
	{\frac{dN}{\sqrt{-g} d^4x}} = \frac{\alpha \tilde E \tilde B}{4 \pi^2 \hbar}\sum_{{l}=1}^\infty\frac1{{l}}\coth\left({l}\,\pi \frac{\tilde B}{\tilde E}\right)\exp{\left({-\frac{{l}\,\pi E_c}{\tilde E}}\right)},
\end{equation}
which explicitly shows an exponential cutoff of the pair creation when the electric field $\tilde {E} $ is {of the order of} $E_c$.

Mathematically, the pair-creation rate becomes exactly zero only in the limit $\tilde E\to 0$. However, the exponential cutoff suggests that we can define the dyadoregion as the region extending bounded from below by the BH horizon and above by the electric field contour \citep{2009PhRvD..79l4002C}
\begin{equation}\label{eq:dyadoregion}
    \tilde{E}(r,\theta) = k\,E_c,
\end{equation}
where $\tilde E$ is given by Eq. (\ref{eq:Eparmodule}), and $k$ is a constant of order unity. The condition (\ref{eq:dyadoregion}) defines an implicit equation for the dyadoregion, $r_d(\theta)$. Equation (\ref{eq:dyadoregion}) with $k=1$ was used to define the dyadosphere of the Reissner-Nordstr\"om BH in \citet{1999A&A...350..334R,2000A&A...359..855R}, while \citet{2009PhRvD..79l4002C} explored different values of $k$ in the dyadotorus of the Kerr-Newman BH. Without loss of generality, we hereafter set $k=1$.

In this paper, we specialize in the case of the Wald solution of a Kerr BH{, of mass $M$ and dimensionless spin parameter $\xi = J/M^2$,} embedded in an external, asymptotically uniform and aligned magnetic field of intensity $B_0$ (see Appendix \ref{app:A}, for details). Figure \ref{fig:Efield} shows, in the $x$-$z$ plane of Kerr-Schild, Cartesian coordinates \citep[see Appendix A.2 in][for details]{2022ApJ...929...56R}, the dyadoregion defined by the contour set by Eq. (\ref{eq:dyadoregion}), using the expression of the electric field {$\tilde E$} given by Eq. (\ref{eq:PAREB2}). In this example, the spin parameter is $\xi = 0.5$ and the magnetic field parameter is {$\beta \equiv B_0/B_c = 200$} ($B_0 \approx 8.8\times 10^{15}$ G){, where $B_c = E_c = m_e^2 c^3/(e \hbar) \approx 4.41\times 10^{13}$ G}. The gray-dashed line delimits the \textit{polar lobes} given by the semi-aperture spherical polar angle $\theta_p \approx \arccos{(\sqrt{3}/3)}\approx 55^\circ$, over which the electric field lines reverse direction. This figure shows that the dyadoregion of the Wald solution is far from being spherically symmetric, which must be accounted for when characterizing the physical quantities related to it, such as its energetics and the pair creation rate.

\begin{figure}
    \centering
    \includegraphics[width=\hsize,clip]{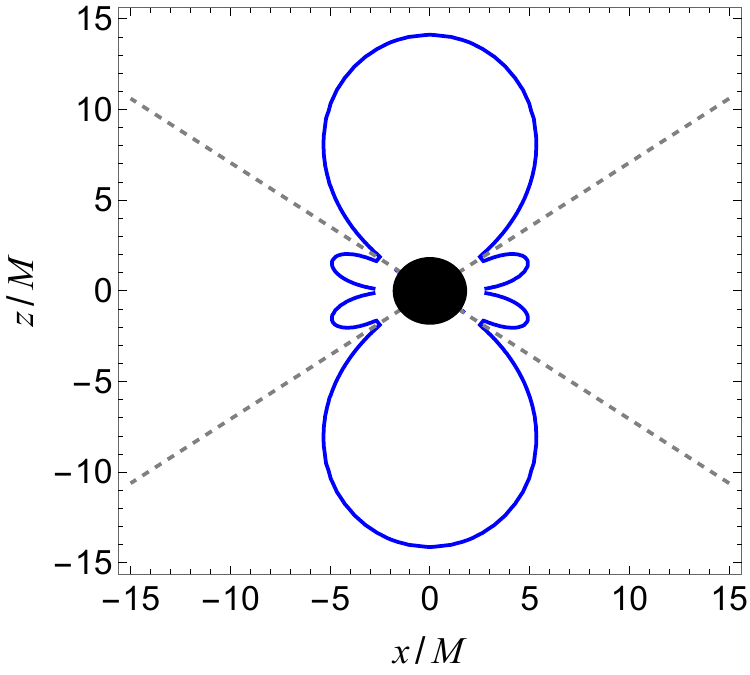}
        \caption{Contour of constant electric field intensity $\tilde E = E_c$ (solid blue curve), in the $x$-$z$ plane of Kerr-Schild, Cartesian coordinates. The black-filled disk is the Kerr BH horizon. In this example, the BH spin parameter is $\xi = 0.5$ and magnetic field strength $\beta = 200$, which corresponds to $B_0 = 8.8\times 10^{15}$ G. The dashed gray lines show the ends of the polar lobes which have a semi-aperture spherical polar angle $\theta_p \approx \arccos{(\sqrt{3}/3)}\approx 55^\circ$.}
    \label{fig:Efield}
\end{figure}

%%%%%%%%%%%%%%%%%%%%%%%%%%%%%%%%%%%%%%%%%%%%%%%%%%%%%%%%%%%%
\subsection{Minimum magnetic field for pair creation}\label{sec:2a}
%%%%%%%%%%%%%%%%%%%%%%%%%%%%%%%%%%%%%%%%%%%%%%%%%%%%%%%%%%%%

We are now ready to estimate the minimum magnetic field for pair creation around the BH. Since {$\tilde E$} decreases with distance, its maximum intensity is at the BH horizon $r = r_+$. Thus, we obtain the lower limit to the magnetic field by requesting that the electric field intensity at the BH horizon be equal to $E_c$. On the polar axis, the minimum magnetic field strength that guarantees this condition is
\begin{equation}\label{eq:Bminpairs}
   \beta_{\rm min} = \frac{B_{0,\rm min}}{B_c} = \frac{2 \bar r_+^2}{\xi (\bar r_+^2 - \xi^2)},
\end{equation}
where we have used that $\tilde{E}(r_+,\theta=0)= \xi\, \beta\, B_c (\bar r_+^2-\xi^2)/(2 \bar r_+^2)$, via Eq. (\ref{eq:EMfieldzamo}), and introduced the dimensionless radial coordinate $\bar r = r/M$, so $\bar r_+ = r_+/M = 1 + \sqrt{1-\xi^2}$. Thus, the value of $B_{0,\rm min}$ depends only on the value of the dimensionless spin parameter, $\xi$. For instance, in the case of a spin $\xi= 0.5$, Eq. (\ref{eq:Bminpairs}) gives a minimum value $B_{0,\rm min} \approx 4.31 B_c \approx 1.90\times 10^{14}$ G. In the example of Fig. \ref{fig:Efield}, we use $\beta = 200$, so $B_0 = 8.8\times 10^{15}$ G, which is above $B_{0, \rm min}$. Indeed, the electric field intensity at the BH horizon is above the critical field, i.e., $\tilde{E} \approx 46.41 E_c$.

%%%%%%%%%%%%%%%%%%%%%%%%%%%%%%%%%%%%%%%%%%%%%%%%%%%%%%%%%%%%
\subsection{Electromagnetic energy in the dyadoregion} \label{sec:2b}
%%%%%%%%%%%%%%%%%%%%%%%%%%%%%%%%%%%%%%%%%%%%%%%%%%%%%%%%%%%%

\begin{figure}
\centering
    \includegraphics[width=\hsize,clip]{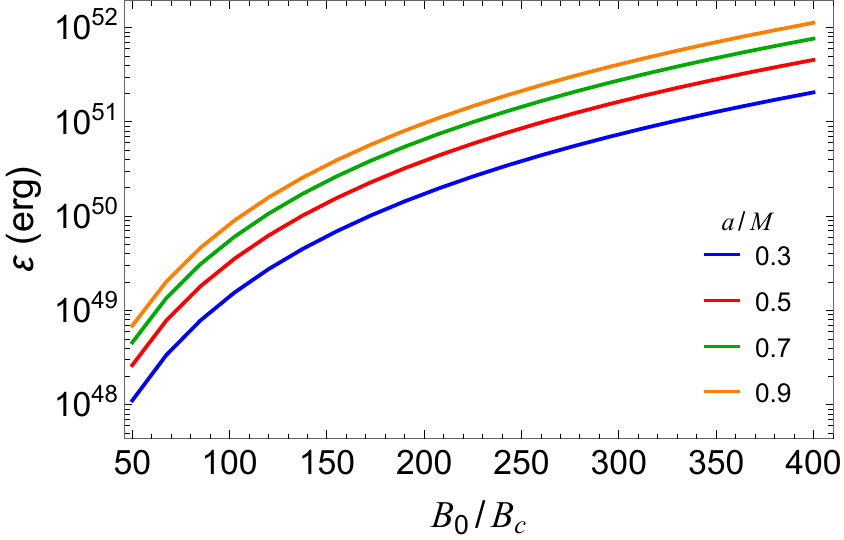}
        \caption{Dyadoregion electromagnetic energy given by Eq. (\ref{eq:energyfinal}), as a function of the magnetic field strength in the range $B_0=(50,400)B_c= (0.22,1.76)\times 10^{16}$ G, for selected values of the BH spin parameter, $a/M = 0.3$ (blue), $0.5$ (red), $0.7$ (green), $0.9$ (orange), and mass $M = 3 M_\odot$.}
    \label{fig:dyadoenergy}
\end{figure}

Figure \ref{fig:dyadoenergy} shows the dyadoregion electromagnetic energy, calculated from Eq. (\ref{eq:energyfinal2}), as a function of $B_0$ and $a/M$. All the magnetic field values are above the minimum value for pair creation given by Eq. (\ref{eq:Bminpairs}).

We can understand the behavior of electromagnetic energy by deriving an approximate, yet accurate, analytic expression for it. First, we note that, as discussed in Appendix \ref{app:C}, the boost is generally weakly relativistic and the electric field $\tilde E$ is well approximated by
\begin{equation}\label{eq:Eapprox}
    \tilde{E}\approx |\hat{E} \cos\theta| \approx \frac{\xi B_0}{\bar r^2} |(3 \cos^2\theta-1)\cos\theta|,
\end{equation}
which is remarkably accurate in the regime of small polar angles. Equation (\ref{eq:Eapprox}) implies that the dyadoregion equation, $r_d(\theta)$, implicitly defined by Eq. (\ref{eq:dyadoregion}), can be approximately described in explicit analytic form by
\begin{equation}\label{eq:rdapprox}
    \bar r_d(\theta) \approx (\xi \beta)^{1/2} |(3 \cos^2\theta-1) \cos\theta|^{1/2}.
\end{equation}
Figure \ref{fig:dyadoenergy} shows the numerical results of the integral (\ref{eq:energyfinal}) evaluated in the dyadoregion. The numerical result is well approximated as follows. Since the electromagnetic energy is dominated by magnetic energy, i.e., $\hat{B} \gg \hat{E}$, Eq. (\ref{eq:energyfinal}) can be approximated by 
\begin{align}
    {\cal E} &\approx \frac{1}{4} \int_0^\pi \int_{r_+}^{r_d(\theta)} \hat{B}^2 \Sigma \sin\theta dr d\theta \approx \frac{B_0^2}{4} \int_0^\pi\int_{r_+}^{r_d(\theta)} r^2 \sin\theta dr d\theta\nonumber\\
    &\approx \frac{B_0^2 M^3}{12} \left(\xi \beta \right)^{3/2} \eta \approx 4.1\times 10^{41} \mu^3\xi^{3/2}\beta^{7/2} \,\,{\rm erg}, \label{eq:energyapprox}
\end{align}
where $\eta \approx 0.78$ and we have only kept the leading order in the rotation parameter, $\xi$. Equation (\ref{eq:energyapprox}) shows that, at fixed spin, the electromagnetic energy increases as $B_0^{7/2}$, which agrees with the numerical result of Fig. \ref{fig:dyadoenergy}. For $\mu=3$, $\xi = 0.9$, and $\beta = 400$, Eq. (\ref{eq:energyapprox}) leads to ${\cal E} = 1.21\times 10^{52}$ erg, and the numerical solution of the integral (\ref{eq:energyfinal}) leads to ${\cal E} = 1.12\times 10^{52}$ erg.

The dyadoregion energy is mostly concentrated in the polar lobes. On the other hand, observational data gives information on the energetics under the assumption of isotropic emission. Therefore,  it is helpful to define a beaming factor that allows estimating the energy stored in the dyadoregion from the knowledge of the isotropic energy release of an observed source, say, $E_{\rm iso}$. The volume of a cone of semi-aperture angle $\theta_b$ is $V_{\rm cone} = (4\pi/3)r^3_{\rm dya}(0)(1-\cos\theta_b)$, where $r_d(0) = M \sqrt{2 \xi \beta}$, so the magnetic energy inside that cone would be ${\cal E}_{\rm cone} = (1/6)B_0^2 M^3 (2 \xi \beta)^{3/2} (1-\cos\theta_b)$. Thus, we define the beaming factor, $f_b$, by requesting the cone's energy equals the dyadoregion energy, i.e., ${\cal E}_{\rm cone}={\cal E}$, which leads to
\begin{equation}\label{eq:thetab}
    f_b \equiv 1-\cos\theta_b = \frac{\eta}{2^{5/2}} \approx 0.14,
\end{equation}
where we have used Eq. (\ref{eq:energyapprox}). Hence, the beaming angle is $\theta_b\approx 30.42^\circ$. Therefore, an approximate value of the dyadoregion energy from the observed isotropic energy is obtained by reducing the latter by the beaming factor $f_b$, i.e., ${\cal E} = f_b E_{\rm iso} \approx 0.14 E_{\rm iso}$, nearly independent on the BH spin. Conversely, we have $E_{\rm iso} \approx 7.14 {\cal E}$. For example, for $\xi = 0.9$, $\mu = 3$, $\beta = 400$, we have ${\cal E} = 1.12\times 10^{52}$ erg (orange curve in Fig. \ref{fig:dyadoenergy}), which implies an approximate isotropic energy equivalent $E_{\rm iso} \approx 8\times 10^{52}$ erg.

%%%%%%%%%%%%%%%%%%%%%%%%%%%%%%%%%%%%%%%%%%%%%%%%%%%%%%%%%%%
%%%%%%%%%%%%%%%%%%%%%%%%%%%%%%%%%%%%%%%%%%%%%%%%%%%%%%%%%%%
\section{Pair-plasma thermodynamical properties}\label{sec:3}
%%%%%%%%%%%%%%%%%%%%%%%%%%%%%%%%%%%%%%%%%%%%%%%%%%%%%%%%%%%
%%%%%%%%%%%%%%%%%%%%%%%%%%%%%%%%%%%%%%%%%%%%%%%%%%%%%%%%%%%

Because the electron-positron-photon plasma thermalizes on rapid timescales as short as $10^{-13}$ s \citep{2007PhRvL..99l5003A,2009PhRvD..79d3008A}, we estimate the initial conditions of the pair plasma thermodynamic properties by assuming they obey Fermi-Dirac statistics. {Thus, the  temperature $T(r,\theta)$ of the plasma is given implicitly by}
\begin{align}\label{eq:nfermidirac}
    n &= \frac{t^3}{\pi^2 \lambda_e^3} \int_0^\infty \frac{y^2}{e^{\sqrt{y^2 + 1/t^2}}+1} dy \approx \frac{3 \zeta(3)}{2\pi^2 \lambda_e^3} t^3,
\end{align}
where $\zeta(s) \approx 1.202$ is the Riemann zeta function, $n$ is the pair density, $y\equiv c\,p/(k_B T)$, $t \equiv k_B T/(m_e c^2)$, being $k_B$ the Boltzmann constant, $p$ the particle momentum, and the last equality is the analytic results in the approximation $k_B T \gg m_e c^2$, which differs at most $10\%$ from the numerical value for $k_B T \sim m_e c^2$. 

{Therefore,} to estimate the plasma temperature, we must estimate the pair density, $n${, measured by a locally Minkowskian observer}. Retaining only the first term of the sum in Eq. (\ref{eq:rate2}), which is sufficiently accurate for our purposes, we can estimate the {invariant} pair creation rate by 
\begin{equation}\label{eq:npairs}
    {\frac{dN}{\sqrt{-g} d^4x} = \frac{dN}{d\tau dV_{\rm prop}}} \approx \frac{\alpha \tilde{E} \tilde{B}}{4\pi^2 \hbar} \coth{\left(\frac{4\pi^2 \tilde{B}}{\tilde{E}}\right)}\exp{\left({-\frac{\pi E_c}{\tilde{E}}}\right)},
\end{equation}
{where we have written the invariant four-volume in terms of the proper time and volume of the locally Minkowskian observer. For the latter, we use the locally non-rotating frame (LNRF), i.e., the \textit{zero angular momentum observer} (ZAMO, \citealp{1972ApJ...178..347B}; see Appendix \ref{app:A}).} Because $\tilde{B}\gg \tilde{E}$, and {taking as $d\tau$} the Compton time $\hbar/(m_e c^2)$, we obtain from Eq. (\ref{eq:npairs}) an accurate approximation of the {local} density of pairs
\begin{equation}\label{eq:npairsapprox}
    n {\equiv \frac{dN}{dV_{\rm prop}}}   \approx \frac{\alpha}{4 \pi^2 m_e c^2} \tilde{E} \tilde{B}\exp{\left({-\frac{\pi E_c}{\tilde{E}}}\right)},
\end{equation}
{where $dV_{\rm prop} = \sqrt{\Sigma A/\Delta}\sin\theta dr d\theta d\phi$ is the ZAMO proper volume (see Appendix \ref{app:A} for the definition of the functions $\Sigma$, $\Delta$, and $A$).} From Eq. (\ref{eq:nfermidirac}), and using Eq. (\ref{eq:npairsapprox}), we obtain the temperature
\begin{equation}\label{eq:tpairs}
    T(r,\theta) = \frac{m_e c^2}{k_B}\frac{\beta^{2/3}}{[6 \zeta(3)^{1/3}]} \left(\frac{\tilde E \tilde B}{B_0^2}\right)^{1/3} \exp{\left({-\frac{\pi E_c}{3\tilde{E}}}\right)}.
\end{equation}

\begin{figure*}
\centering
\includegraphics[width=0.49\hsize,clip]{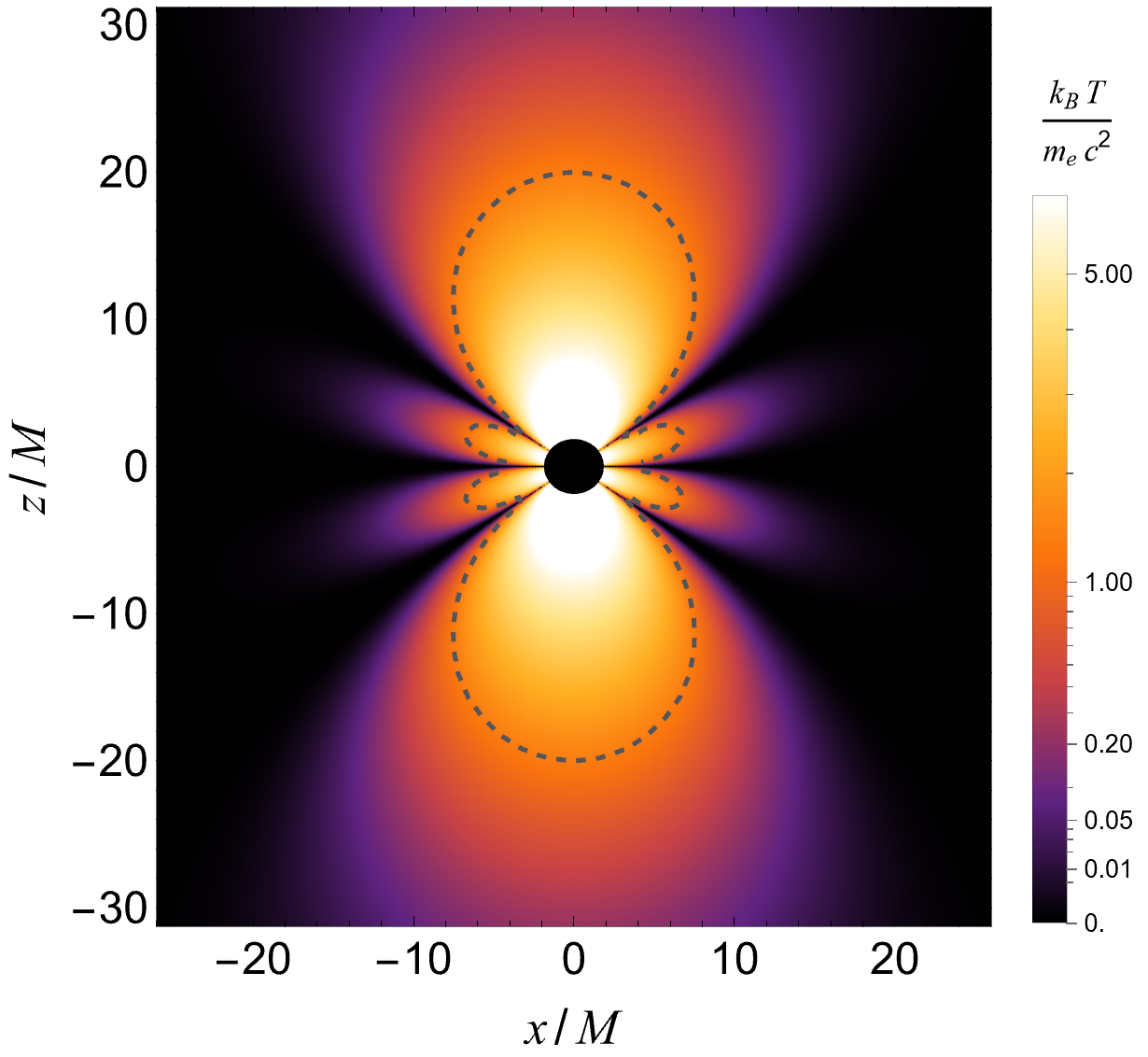}\includegraphics[width=0.49\hsize,clip]{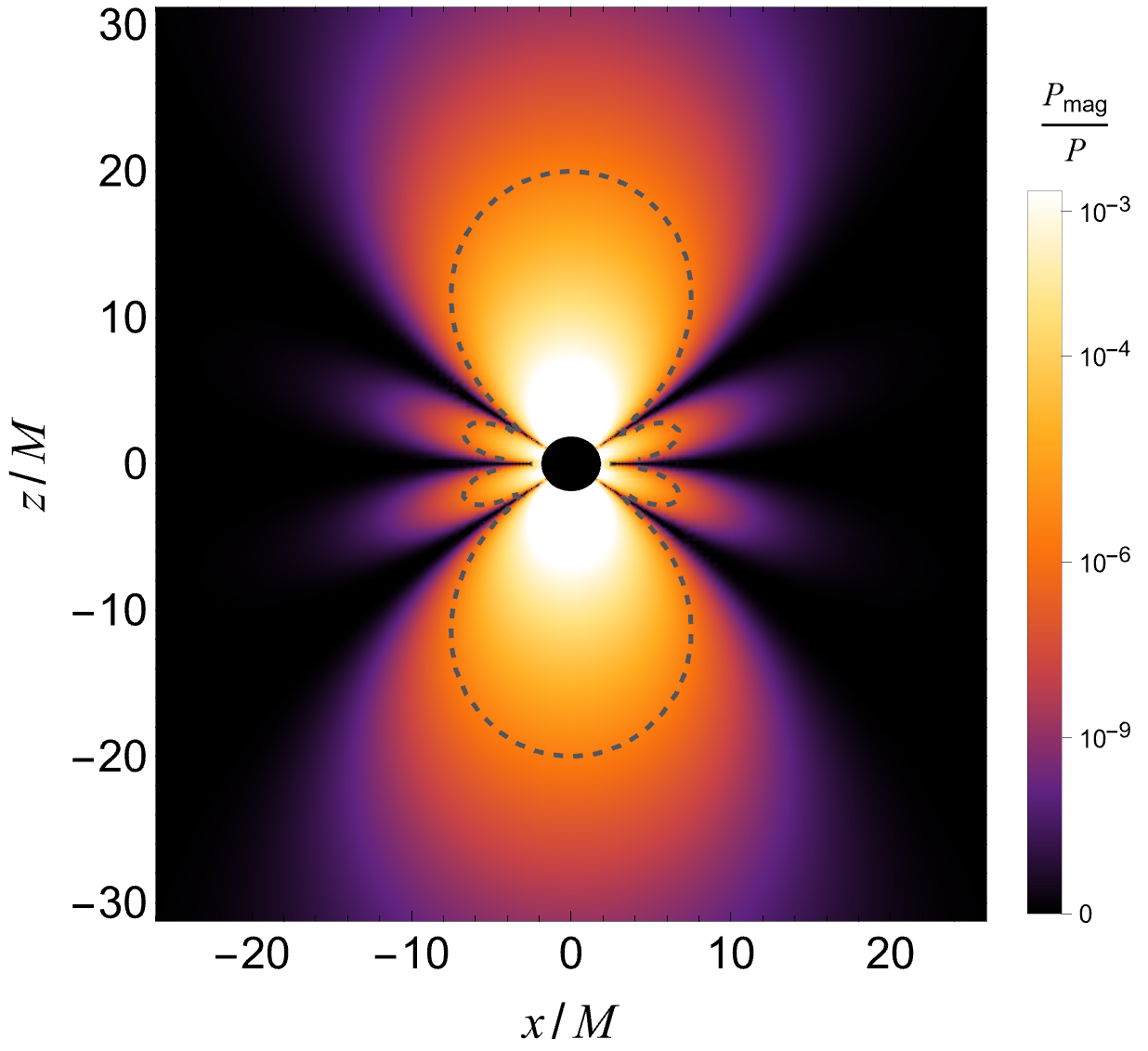}\\
    \includegraphics[width=0.476\hsize,clip]{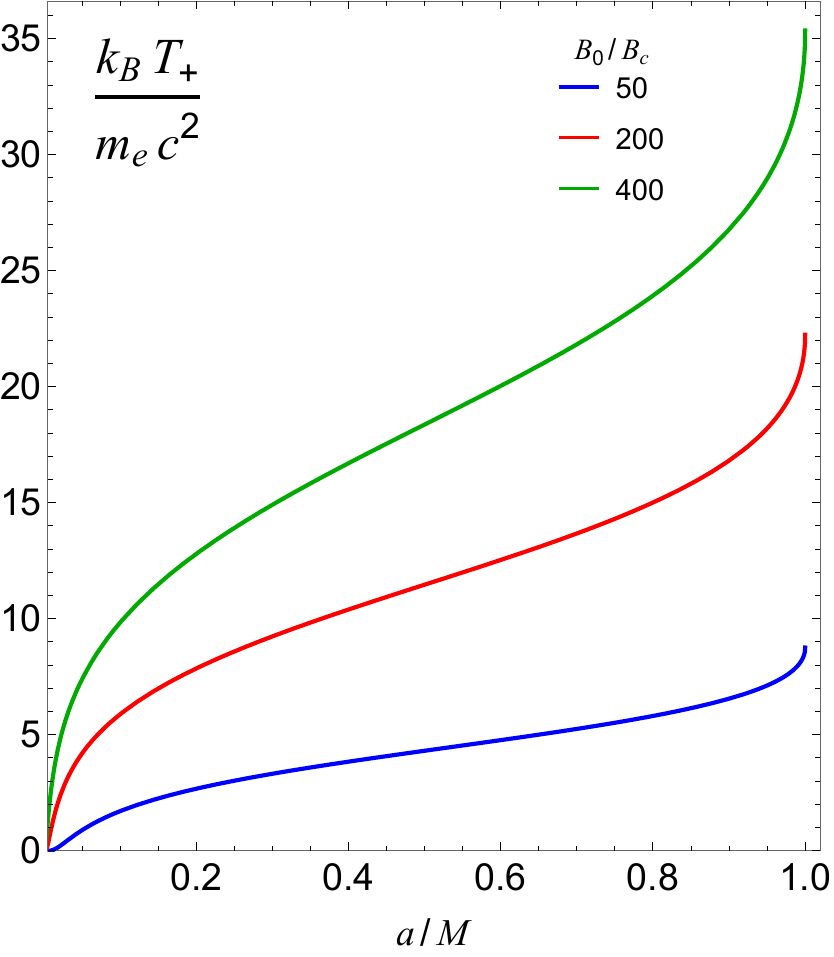}\includegraphics[width=0.49\hsize,clip]{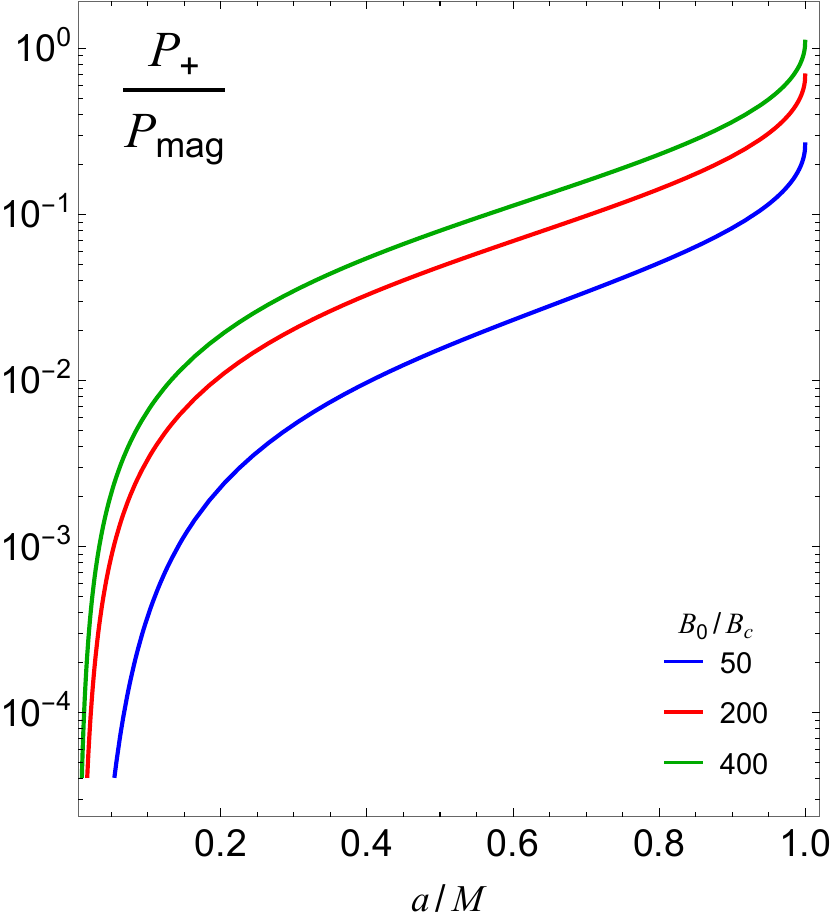}
        \caption{Upper left: plasma temperature $k_B T/(m_e c^2)$ around the Kerr BH of spin parameter $\xi = 0.5$ and magnetic field strength parameter $\beta = 400$. The dark-gray dashed contour is the dyadoregion radius given by the condition $\tilde{E}=E_c$. Upper right: plasma parameter $P/P_{\rm mag}$ for the same parameters as the upper right plot. Lower left: plasma temperature at the horizon, $k_B T_+/(m_e c^2)$, as a function of the BH spin $\xi=a/M$, for selected values of the magnetic field $\beta=50$ (blue), $200$ (red), and $400$ (green). Lower right: Plasma parameter at the horizon, $P_+/P_{\rm mag}$, for the same parameter as the lower left plot. {The temperature and pressure of the plasma are given in Eqs. (\ref{eq:tpairs}) and (\ref{eq:plasmaEOSP}), respectively}.
        }
    \label{fig:plasma}
\end{figure*}

Inside the dyadoregion, the energy density and pressure of the $e^+e^-\gamma$ plasma are given by
\begin{align}
    \epsilon &=a T^4\left(1 + \frac{30}{\pi^4} \int_0^\infty \frac{\sqrt{y^2+1/t^2} y^2}{e^{\sqrt{y^2 + 1/t^2}}+1} dy \right) \approx \frac{11}{4} a T^4,\label{eq:plasmaEOS}\\
    P &= \frac{1}{3}a T^4\left(1 + \frac{30}{\pi^4} \int_0^\infty \frac{(y^2+1/t^2)^{-1/2} y^2}{e^{\sqrt{y^2 + 1/t^2}}+1} dy \right) \approx \frac{\epsilon}{3},\label{eq:plasmaEOSP}
\end{align}
where $a = 4 \sigma/c$, being $\sigma = 2\pi^5 k_B^4/(15 h^3 c^2)$ the Stefan-Boltzmann constant. 

Figure \ref{fig:plasma} (left upper panel) shows the temperature of the pairs given by Eq. (\ref{eq:tpairs}) in the case of a BH spin $\xi = 0.5$ and magnetic field strength parameter $\beta = 400$. At the dyadoregion border, $r=r_d$, using Eq. (\ref{eq:rdapprox}), we obtain the temperature $k_B T_{\rm dya}\equiv k_B T(r_d,\theta) = m_e c^2 [6 \zeta(3)]^{-1/3} \beta^{1/3}e^{-\pi/3}$, so $k_B T_{\rm dya}\approx 1.3 m_e c^2$ for these parameters. Near the BH horizon, at the pole ($\theta=0$), the temperature is $k_B T_+ = k_B T(r_+,0) \approx 18.4\,m_e c^2$. The value of the temperature in the region near the BH (at $r<10 M$) is not well resolved by the left-panel plot, so in the upper right panel, we show the temperature at the horizon, $k_B T_+/(m_e c^2)$, as a function of the BH spin, for selected $B_0$ strengths.

For the above parameters, we have at the BH horizon, at the pole, $\epsilon_+ \equiv \epsilon (r_+,0) \approx 2.9\times 10^{30}$ erg cm$^{-3}$ and $P_+ \equiv P(r_+,0) \approx 9.8 \times 10^{29}$ erg cm$^{-3}${, where the energy density and pressure are given in Eqs. (\ref{eq:plasmaEOS}) and (\ref{eq:plasmaEOSP}), respectively}. Interestingly, $\epsilon_+$ and $P_+$ are much smaller than the magnetic pressure or energy density, $P_{\rm mag} = B_0^2/(8\pi) = \beta^2 B_c^2/(8 \pi)\approx 1.2\times 10^{31}$ erg cm$^{-3}$. Figure \ref{fig:plasma} (lower left) shows the plasma parameter, $P/P_{\rm mag}$, in the case of a BH spin $\xi =0.5$ and magnetic field $\beta = 400$. The value of the ratio in the region near the BH (at $r<10 M$) is not well resolved in the plot, so in the lower right panel of the figure, we complementarily show the plasma parameter at the horizon, $P_+/P_{\rm mag}$, as a function of the BH spin and for various magnetic field strengths. Because $T \propto \beta^{2/3}$, the plasma over the magnetic field pressure ratio is $\propto T^4/B_0^2 \propto \beta^{8/3}/\beta^2 = \beta^{2/3}$, so the ratio becomes lower for smaller $B_0$. This result suggests the possibility of evolving the pair plasma in a magnetized medium via ideal magnetohydrodynamics (MHD), with magnetic energy being converted into the kinetic energy of the pair plasma, leading to ultrarelativistic motion.

%%%%%%%%%%%%%%%%%%%%%%%%%%%%%%%%%%%%%%%%%%%%%%%%%%%%%%%%%%%
%%%%%%%%%%%%%%%%%%%%%%%%%%%%%%%%%%%%%%%%%%%%%%%%%%%%%%%%%%%
\section{Discussion and conclusions}\label{sec:4}
%%%%%%%%%%%%%%%%%%%%%%%%%%%%%%%%%%%%%%%%%%%%%%%%%%%%%%%%%%%
%%%%%%%%%%%%%%%%%%%%%%%%%%%%%%%%%%%%%%%%%%%%%%%%%%%%%%%%%%%

We have comprehensively characterized, in an invariant fashion, the dyadoregion around a Kerr BH in the presence of an external, test, asymptotically uniform magnetic field aligned to the BH spin (the Wald solution). We calculated the size, morphology, and thermodynamic properties of the dyadoregion as a function of the BH spin and magnetic field strength. We have shown that only external magnetic fields with a strength above a few $10^{14}$ G can induce an overcritical electric field that creates $e^+e^-$ pairs.

Stellar-mass BHs formed from the collapse of an NS might be surrounded by a magnetic field that could be nearly uniform within a few horizon radii, and with strengths well above $B_c = 4.4\times 10^{14}$ G, as a result of the rapid increase in field strength during the collapse. Numerical simulations of rotating magnetized collapse into a BH show that magnetic field amplification can be even greater than predicted by magnetic flux conservation \citep{2013PhRvD..88d4020D,2017MNRAS.469L..31N,2018ApJ...864..117M}. However, a uniform field strength of, e.g., $10^{16}$ G, extending over a region beyond ten horizon radii, would imply an extraordinary electromagnetic energy density likely unsustainable in realistic astrophysical scenarios. Interactions with the surrounding plasma, relativistic effects, and magnetic reconnection might cause the field to deviate significantly from homogeneity at those distances, or the BH and the surrounding material could be unable to sustain those strong magnetic fields. More complex simulations incorporating general relativistic magnetohydrodynamics are required to model those environments around the BH. However, we have shown that most $e^+e^-$ pairs are produced near the horizon, so the thermodynamic properties of the pair-photon plasma are dominated by their near-horizon values. Thus, the simplified treatment based on the Wald solution can provide an accurate theoretical insight.

The formulation presented herein establishes a foundation for understanding the initial conditions of the dynamics of the $e^+e^-\gamma$ plasma in extreme astrophysical environments associated with high-energy transients like GRBs. The dynamics and final transparency of the plasma depend on the amount of baryonic matter engulfed during its expansion. Numerical simulations considering the axially symmetric morphology and the initial conditions of the thermodynamic properties of the plasma presented here are needed to analyze the temporal and spectral data of the UPE in GRBs. For instance, since the electromagnetic energy deposited in the pairs is stored in axially symmetric polar lobes, the baryon load of the expanding plasma could be lower than spherically symmetric situations  \citep[see, e.g.][]{1999A&A...350..334R,2000A&A...359..855R,2018ApJ...852...53R,2021PhRvD.104f3043M,2022EPJC...82..778R,2023Symm...15..412C}. The detailed dynamics of the axially symmetric expanding pair plasma, considering the role of internal anisotropy pressure, could provide valuable insights into the possible collimation and the mechanisms driving highly relativistic jetted emission. 

We have obtained the distribution of the $e^+e^-\gamma$ plasma temperature in the dyadoregion (upper left panel of Fig. \ref{fig:plasma}). For a given magnetic field strength, the lower the BH spin, the lower the temperature (upper right panel of Fig. \ref{fig:plasma}). This result suggests that the energy per pair should decrease with the rotation rate. If the photon energy at transparency inherits the hard-to-soft evolution of the plasma energy, a decreasing peak photon energy could be observed in time-resolved spectral analyses of the emission when angular momentum losses are at work.

We have shown that, at the beginning, the magnetic pressure is higher than the $e^+e^-\gamma$ plasma pressure (see Fig. \ref{fig:plasma}), suggesting the survival of the magnetic field during the plasma expansion. The consequences of these initial conditions for the subsequent plasma dynamics, taking into account the properties of the surrounding medium, also warrant further investigation via relativistic magnetohydrodynamic simulations. 

Further, the presence of instabilities within the expanding plasma instabilities may give rise to complex, three-dimensional self-organized criticality and self-similar behavior, which could manifest during the prompt emission phase of GRBs \citep[see, e.g.][]{2021FrPhy..1614501L,2023ApJS..265...56L}. A deeper understanding of these instabilities may provide a novel perspective on GRB prompt emission variability and structural complexity.

The treatment presented here can be extended to the analysis of QED pair creation in the surroundings of a highly magnetized, fast-rotating NS approaching the critical mass for BH formation. The intensity of the electric field induced by a co-rotating magnetic field, $E \sim (v/c)B = (\Omega R/c) B \approx 0.2 (R_6/P_{\rm ms}) E_c (B/B_c)$, where $R_6$ and $P_{\rm ms}$ are the NS radius and rotation period in units of $10^6$ cm and milliseconds, can exceed $E_c$ for magnetic field strengths above a few $B_c$. Although the Wald solution offers insight into such an astrophysical situation, its accurate modeling must be done within the appropriate solution of the Einstein-Maxwell equations, which is left for forthcoming studies.

Numerical simulations \citep[also three-dimensional, e.g.,][]{2006PhRvL..96c1101D,2006PhRvL..96c1102S,2006PhRvD..73j4015D,2007CQGra..24S.207S,2008PhRvD..77d4001S,2011ApJ...732L...6R} show that the central remnant could be initially a fast-rotating merged core with a strong electromagnetic field, then collapsing into a rotating BH. Indeed, an analogous physical situation to the one here analyzed may occur around the merged core of NS-NS mergers that lead to a Kerr BH \citep{2026JHEAp..5000464R}.

\section*{Acknowledgments}

C.C. wishes to acknowledge the Gruppo Nazionale di Fisica Matematica GNFM-INdAM.

\appendix

%%%%%%%%%%%%%%%%%%%%%%%%%%%%%%%%%%%%%%%%%%%%%%%%%%%%%%%%
%%%%%%%%%%%%%%%%%%%%%%%%%%%%%%%%%%%%%%%%%%%%%%%%%%%%%%%%
\section{Electromagnetic field of the Wald solution in the locally non-rotating frame}\label{app:A}
%%%%%%%%%%%%%%%%%%%%%%%%%%%%%%%%%%%%%%%%%%%%%%%%%%%%%%%%
%%%%%%%%%%%%%%%%%%%%%%%%%%%%%%%%%%%%%%%%%%%%%%%%%%%%%%%%

In this appendix, we derive the electric and magnetic fields associated with the Wald solution \citep{1974PhRvD..10.1680W}, as measured by the locally non-rotating frame (LNRF), i.e., the \textit{zero angular momentum observer} (ZAMO, \citealp{1972ApJ...178..347B}){, say $\bf\hat E$ and $\bf\hat B$}. {As shown by Eq. (\ref{eq:rate2}) in Section \ref{sec:2}, the relevant quantities for calculating the pair creation rate are the moduli of the electric and magnetic fields in the frame where they are parallel to each other, say $\tilde E$ and $\tilde B$. The reason we introduce the fields $\bf\hat E$ and $\bf\hat B$ measured by the ZAMO is twofold. First, we can obtain the fields $\bf\tilde E$ and $\bf\tilde B$ by applying a Lorentz boost to $\bf\hat E$ and $\bf\hat B$ (see Appendix \ref{app:C}). Second, the fields $\bf\hat E$ and $\bf\hat B$ naturally appear in the expression of electromagnetic energy (see Appendix \ref{app:B} for more details).}

We start with the Kerr BH spacetime metric
\begin{align}\label{KmetricNP}
    ds^2 &= - \left(1-\frac{2 M r}{\Sigma}\right) dt^2 - \frac{4 a M r}{\Sigma} \sin^2\theta\, dt\, d\phi + \frac{\Sigma}{\Delta} dr^2 \nonumber \\
    &+ \Sigma\,d\theta^2 + \frac{A}{\Sigma}\sin^2\theta\,d\phi^2,
\end{align}
where $\Sigma=r^2+a^2\cos^2\theta$, $\Delta=r^2-2 M r+ a^2$, $A = (r^2+a^2)^2-\Delta a^2 \sin^2\theta$, being $M$ and $a=J/M$, respectively, the BH mass and angular momentum per unit mass.

The basis vectors carried by ZAMO are $e_{\hat a} = e_{\hat a}^{\hphantom{\hat a}{\mu}} e_\mu$, where $e_\mu$ are the basis vectors of the coordinate frame. Thus, the transformation between the coordinate frame and the LNRF is
\begin{equation}
    [e_{\hat a}^{\hphantom{\hat a}{\mu}}] = \left(
    \begin{array}{cccc}
        \Gamma & 0 & 0 & \Gamma \omega \\
         0 & \sqrt{\frac{\Delta}{\Sigma}} & 0 & 0 \\
         0 & 0 & \frac{1}{\sqrt{\Sigma}} & 0 \\
         0 & 0 & 0 & \sqrt{\frac{\Sigma}{A}}\frac{1}{\sin\theta}
    \end{array}
    \right),
\end{equation}
with $\Gamma = \sqrt{A/(\Sigma \Delta)}$ and $\omega = -g_{03}/g_{33} = 2 M a r/A$. The ZAMO four-velocity components, $u^\mu_{(Z)}$, are defined through $e_{\hat{0}}  = u^\mu_{(Z)}e_\mu$, so $u^\mu_{(Z)} =\Gamma(1,0,0,\omega)$.

The electric and magnetic field components in the LNRF are $E_{\hat{i}} = e_{\hat i}^{\hphantom{\hat i}{j}} E_j$ and $
B_{\hat{i}} = e_{\hat i}^{\hphantom{\hat i}{j}} B_j$, where $E_j = F_{j \nu} u^\nu_{(Z)}$ and $B_j = \tilde{F}_{j \nu} u^\nu_{(Z)}$. The electromagnetic field tensor is $F_{\mu\nu} = \partial_\mu A_\nu - \partial_\nu A_\mu$ and its dual $\sqrt{-g} \tilde{F}^{\mu \nu} = (1/2) \epsilon^{\mu \nu \alpha \beta}F_{\alpha \beta}$, where $\epsilon^{\mu \nu \alpha \beta}$ is the Levi-Civita symbol{, and $g=- \Sigma^2 \sin^2\theta$ the metric determinant}. 

The electromagnetic four-potential of the Wald solution is \citep{1974PhRvD..10.1680W} 
\begin{equation}\label{eq:Amu}
    A_\mu =(B_0/2)\,\psi_\mu + a \,B_0 \xi_\mu,
\end{equation}
where $\xi^\mu= \delta^\mu_{\hphantom{\mu}{0}}$ and $\psi^\mu=\delta^\mu_{\hphantom{\mu}{3}}$ are the time-like and space-like Killing vectors of the Kerr spacetime. Thus, $A_\mu = (A_0,0,0,A_3)$ with
\begin{subequations}\label{eq:potential}
    \begin{align}
    A_0 &= -a B_0 \Bigg[ 1 - \frac{M r}{\Sigma} (1+\cos^2\theta) \Bigg],\\
    A_3 &= \frac{1}{2} B_0\sin^2\theta  \Bigg[ r^2 + a^2 - \frac{2 M r a^2}{\Sigma} (1+\cos^2\theta) \Bigg].
\end{align}
\end{subequations}
The electromagnetic (Maxwell) field tensor is $F_{\alpha \beta} = \partial_\alpha A_\beta - \partial_\beta A_\alpha$, which for the electromagnetic four-potential (\ref{eq:potential}) has the non-vanishing components
\begin{subequations}\label{eq:Fab}
    \begin{align}
    F_{01} &= \frac{a B_0 M}{\Sigma^2}(r^2-a^2 \cos^2\theta)(1+\cos^2\theta),\\
    F_{02} &= \frac{2 a B_0 M r \sin\theta\cos\theta}{\Sigma^2}(r^2-a^2),\\
    F_{13} &=B_0\sin^2\theta \left[r+ \frac{M a^2 (r^2-a^2 \cos^2\theta)(1+\cos^2\theta)}{\Sigma^2}\right],\\
    F_{23} &= \frac{B_0 \sin\theta\cos\theta}{\Sigma^2} \left[ \Sigma^2 (r^2 + a^2) -2 M a^2 r \Sigma (1+\cos^2\theta) \right.\nonumber \\
    & \left.+ 2 M a^2 r (r^2-a^2)\sin^2\theta \right].
\end{align}
\end{subequations}
Thus, the electric and magnetic field components in the LNRF are given by
\begin{subequations}
    \begin{align}
        E_{\hat 1} = e_{\hat 1}^{\hphantom{\hat 1}{1}} E_1 = \sqrt{\frac{\Delta}{\Sigma}}E_1,\quad        E_{\hat 2} = e_{\hat 2}^{\hphantom{\hat 2}{2}} E_2= \sqrt{\frac{1}{\Sigma}}E_2,\\
        B_{\hat 1} = e_{\hat 1}^{\hphantom{\hat 1}{1}} B_1= \sqrt{\frac{\Delta}{\Sigma}}B_1,\quad
        B_{\hat 2} = e_{\hat 2}^{\hphantom{\hat 2}{2}} B_2 = \sqrt{\frac{1}{\Sigma}}B_2,
    \end{align}
\end{subequations}
where
\begin{subequations}\label{eq:EB}
    \begin{align}
    E_1 &=u^0_{(Z)} (F_{10} + \omega F_{13}),\quad
    E_2 = u^0_{(Z)} (F_{20} + \omega F_{23}),\\
    B_1 &= \sqrt{\frac{g_{11}}{g_{22}g_{33}}}F_{23},\quad
    B_2 = -\sqrt{\frac{g_{22}}{g_{11}g_{33}}}F_{13}.
\end{align}
\end{subequations}
From the above, we obtain the explicit form of the electromagnetic field components in the LNRF \citep{2022ApJ...929...56R,2023EPJC...83..960R,2024EPJC...84.1166R}
\begin{subequations}\label{eq:EMfieldzamo}
\begin{align}
E_{\hat{1}} &= -\frac{B_0 a M}{\Sigma^2 \sqrt{A}} \Bigg[(r^2 + a^2)(r^2-a^2\cos^2\theta)(1+\cos^2\theta) \nonumber \\
&- 2 r^2 \sin^2\theta\,\Sigma\Bigg],\\
E_{\hat{2}} &= B_0 a M \sqrt{\frac{\Delta}{A}}\frac{2 r a^2 \sin\theta \cos\theta (1+\cos^2\theta)} {\Sigma^2},\\
B_{\hat{1}} &= -\frac{B_0 \cos\theta}{\Sigma^2 \sqrt{A}} \{2 M a^2 r [\Sigma (1 + \cos^2\theta)-(r^2-a^2) \sin^2\theta] \nonumber \\
&-(r^2+a^2)\Sigma^2\},\\
B_{\hat{2}} &= -\frac{B_0 \sin\theta}{\Sigma^2}\sqrt{\frac{\Delta}{A}} [M a^2 (r^2-a^2\cos^2\theta)(1+\cos^2\theta) \nonumber \\
&+ r \Sigma^2].
\end{align}
\end{subequations}
The intensity of the above electric and magnetic fields is
\begin{subequations}\label{eq:EBnorms}
    \begin{align}
    \hat{E} \equiv \sqrt{E_{\hat{i}}E^{\hat{i}}} = \sqrt{E^2_{\hat{1}} + E^2_{\hat{2}}},\quad \hat{B} \equiv \sqrt{B_{\hat{i}}B^{\hat{i}}} = \sqrt{B^2_{\hat{1}} + B^2_{\hat{2}}},
\end{align}
\end{subequations}
where we have used $E^{\hat{i}} = E_{\hat{i}}$ and $B^{\hat{i}} = B_{\hat{i}}$.

%%%%%%%%%%%%%%%%%%%%%%%%%%%%%%%%%%%%%%%%%%%%%%%%%%%%%%%%%%%
%%%%%%%%%%%%%%%%%%%%%%%%%%%%%%%%%%%%%%%%%%%%%%%%%%%%%%%%%%%
\section{Electromagnetic energy budget}\label{app:B}
%%%%%%%%%%%%%%%%%%%%%%%%%%%%%%%%%%%%%%%%%%%%%%%%%%%%%%%%%%%
%%%%%%%%%%%%%%%%%%%%%%%%%%%%%%%%%%%%%%%%%%%%%%%%%%%%%%%%%%%

This appendix presents the expression for evaluating the electromagnetic energy of the dyadoregion. For this task, we evaluate the electromagnetic energy stored in a $t$-constant hypersurface ${\cal S}$ which is given by
\begin{equation}\label{eq:energy}
    {\cal E} = \int_{\cal S} T_{\alpha \beta} \xi^\alpha d {\cal S}^\beta,
\end{equation}
where $T_{\alpha \beta}$ is the electromagnetic energy-momentum tensor
\begin{equation}\label{eq:Tmunu}
    T_{\alpha \beta} = -\frac{1}{4\pi} \left(F_{\alpha \mu}F^{\mu}_{\,\beta} +\frac{1}{4}g_{\alpha \beta}{\cal F}\right),
\end{equation}
being ${\cal F}\equiv F_{\mu \nu}F^{\mu \nu}$, and $d{\cal S}^\beta = n^\beta d{\cal S}$ is the surface element vector with $n$ the unit time-like normal to ${\cal S}$. The equation of the $t$-constant hypersurface is $\Phi = t - T = 0$, for an arbitrary coordinate time $T$. Thus, we have $n^\beta = \Phi^{,\beta}/\sqrt{-\Phi^{,\mu}\Phi_{,\mu}} = \Gamma (\xi^\beta + \omega \psi^\beta)$. Notice that $n^\beta= u^\beta_{(Z)}$, the ZAMO four-velocity. Thus, Eq. (\ref{eq:energy}) becomes
\begin{equation}\label{eq:energy2}
    {\cal E} = \int_{\cal S} (T_{00} + \omega T_{03}) \Gamma d {\cal S},
\end{equation}
where $d {\cal S} = \sqrt{g_{11}g_{22}g_{33}} dr d\theta d\phi = \sqrt{\Sigma A/\Delta} \sin\theta dr d\theta d\phi$. Introducing the energy-momentum tensor (\ref{eq:Tmunu}) into Eq. (\ref{eq:energy2}), we can express the electromagnetic energy as
\begin{equation}\label{eq:energyfinal}
    {\cal E} = 2 \pi\int_0^\pi\int_{r_+}^{r(\theta)} \left(U_{\rm em} + S_{\hat 3} \frac{A\sin\theta}{\Sigma\sqrt{\Delta} }\right)\Sigma\sin\theta dr d\theta,
\end{equation}
where $U_{\rm em} = (\hat{E}^2 + \hat{B}^2)/(8\pi)$ is the electromagnetic energy density, with $\hat{E}^2 \equiv E_{\hat{i}}E^{\hat{i}}$ and $\hat{B}^2 \equiv B_{\hat{i}}B^{\hat{i}}$ [see Eq. (\ref{eq:EBnorms})], and $S_{\hat 3} = |{\bf \hat E}\times {\bf \hat B}|/(4\pi) = (E_{\hat{1}} B_{\hat{2}} - E_{\hat{2}} B_{\hat{1}})/(4 \pi)$ is the Poynting vector, being $E_{\hat{i}}$ and $B_{\hat{i}}$ the electric and magnetic field components measured in the LNRF, which for the Wald solution are given in Eqs. (\ref{eq:EMfieldzamo}). The integral is carried out within the surface $r(\theta)$, which delimits the region under consideration (e.g., the dyadoregion).

We can get a glimpse of the energetics by considering a surface of constant radius, $r(\theta) = R$, and neglecting the corrections of rotation ($\Sigma \to r^2$, $A \to r^4$, $\Delta = r^2-2 M r$), for which Eq. (\ref{eq:energyfinal}) reduces to the well-known result
\begin{equation}\label{eq:energyspherical}
    {\cal E} \approx \frac{4 \pi}{3} (R^3-r_+^3) U_{\rm em} = \frac{R^3-r_+^3}{6}\left(\hat{E}^2 + \hat{B}^2\right),
\end{equation}
assuming uniform and parallel (in the LNRF) fields. For instance, if $\hat E \ll \hat B \sim 10^{15}$ G, and $R \sim 10^7$ cm, Eq. (\ref{eq:energyspherical}) leads to ${\cal E} = 1.67 \times 10^{50}$ erg. The dyadoregion energetics, including the effects of rotation, can increase by about one order of magnitude, and the magnetic field of the dyadoregion can be larger, leading to electromagnetic energies of even a few $10^{52}$ erg. 

%%%%%%%%%%%%%%%%%%%%%%%%%%%%%%%%%%%%%%%%%%%%%%%%%%%%%%%%
%%%%%%%%%%%%%%%%%%%%%%%%%%%%%%%%%%%%%%%%%%%%%%%%%%%%%%%%
\section{Parallel electric and magnetic fields of the Wald solution}\label{app:C}
%%%%%%%%%%%%%%%%%%%%%%%%%%%%%%%%%%%%%%%%%%%%%%%%%%%%%%%%
%%%%%%%%%%%%%%%%%%%%%%%%%%%%%%%%%%%%%%%%%%%%%%%%%%%%%%%%

This appendix derives the intensity of the electric and magnetic fields of the Wald solution measured in a frame where they are parallel, {$\tilde E$ and $\tilde B$. First, we use the Newman-Penrose (NP) formalism \citep{1962JMP.....3..566N} (see also \citealp{1998mtbh.book.....C}), which allows us to calculate the moduli $\tilde E$ and $\tilde B$ by-passing the need to determine the specific reference frame.}

{In this appendix}, we use the signature $(+,-,-,-)$ of the Kerr metric (\ref{KmetricNP}) to follow the original treatment of the NP formalism. We refer to \citet{2009PhRvD..79l4002C} for the NP treatment of the Kerr-Newman fields. Using the Kinnersley NP principal tetrad \citep{1969JMP....10.1195K}
\begin{align}
\label{kinns}
l^{\mu}&=\frac{1}{\Delta}[r^2+a^2,\Delta,0,a]\ , \quad 
n^{\mu}=\frac{1}{2\Sigma}[r^2+a^2,-\Delta,0,a]\ ,\nonumber \\
m^{\mu}&=\frac{1}{\sqrt{2}(r+ia\cos\theta)}\,\left[{ia}\,{\sin\theta},0,1,\frac{i}{\sin\theta}\right]\ ,
\end{align}
one gets the nonvanishing spin coefficients 
\begin{align}
\rho&=-\frac{1}{r-ia\cos\theta}, \quad
\tau=-\frac{ia}{\sqrt{2}}\rho\bar \rho\sin\theta, \quad
\beta=-\frac{\bar \rho}{2\sqrt{2}}\cot\theta, \nonumber \\
\pi&=\frac{ia}{\sqrt{2}}\rho^2\sin\theta, \quad
\mu=\frac12\rho^2\bar \rho\Delta, \quad
\gamma=\mu+\frac12\rho\bar \rho(r-M),\nonumber \\ 
\alpha&=\pi-\bar \beta,
\end{align}
where the bar stands for complex conjugation. The only nonvanishing Weyl scalar is $\psi_2=M\rho^3$, which shows that the Kerr solution is of Petrov type D. Once we introduce the Wald electromagnetic four-potential (\ref{eq:Amu}), we can build the Maxwell tensor (\ref{eq:Fab}), project it on the NP Kinnersley tetrad and construct the three complex scalars that represent the electromagnetic field
\begin{equation}
\phi_0=F_{lm},\quad \phi_2=F_{ \bar m n},\quad \phi_1=\frac12(F_{ln}+F_{ \bar m m}),
\end{equation}
which, in the present case, leads to
\begin{eqnarray}
	\label{maxscals}
	\phi_0&=&-\frac{i}{\sqrt{2}}B_0\sin\theta \,,\quad 	\phi_2 = \frac{i}{2\sqrt{2}}B_0\Delta\rho^2\sin\theta \\ 
	\phi_1&=& B_0 a M\rho^2-\frac{B_0(i\cos\theta+aM\rho^2\sin^2\theta-a\rho\sin^2\theta)}{2}\,. \nonumber
\end{eqnarray}
From these scalars, we can compute the electromagnetic invariants (star stands for duality operation) 
\begin{eqnarray}
\label{invariants}
{\mathcal F}&\equiv&\frac14F_{\mu\nu}F^{\mu\nu}=\frac12({\bf B}^2-{\bf E}^2)=2{\rm Re}(\phi_0\phi_2-\phi_1^2)\ , \nonumber\\
{\mathcal G}&\equiv&\frac14F_{\mu\nu}{}^*F^{\mu\nu}={\bf E}\cdot {\bf B}=-2{\rm Im}(\phi_0\phi_2-\phi_1^2)\ ,
\end{eqnarray}
where ${\bf E}$ and ${\bf B}$ represent the electric and magnetic fields respectively. Following \citet{1975PhRvL..35..463D}, we want now to calculate the electric and magnetic fields as measured by an observer who sees them parallel, ${\bf \tilde E}$ and ${\bf \tilde B}$, in analogy with the situation of the Carter observer in the Kerr-Newman spacetime \citep{1968CMaPh..10..280C}. The relations just derived become
\begin{equation}
\label{invariants2}
\tilde B^2-\tilde E^2=4{\rm Re}(\phi_0\phi_2-\phi_1^2)\ , \quad
\tilde E \tilde B=-2{\rm Im}(\phi_0\phi_2-\phi_1^2)\,.
\end{equation}
where $\tilde E = |{\bf \tilde E}|$ and $\tilde B = |{\bf \tilde B}|$. From Eqs. (\ref{invariants}) and  (\ref{invariants2}), one can obtain  expressions for $\tilde E$ and $\tilde B$ in terms of the invariants
\begin{equation}\label{eq:PAREB}
	\tilde E=[({\mathcal F}^2+{\mathcal G}^2)^{1/2}-{\mathcal F}]^{1/2},\quad \tilde B=[({\mathcal F}^2+{\mathcal G}^2)^{1/2}+{\mathcal F}]^{1/2}.
\end{equation}

Then, by {substituting} Eq. (\ref{maxscals}) into Eq. (\ref{invariants}), and the latter into Eq. (\ref{eq:PAREB}), one can obtain after some straightfoward but tedious algebra, explicit expressions of $\tilde E$ and $\tilde B$ as a function of the parameters $B_0$, $a$, $M$, and the coordinates $r$ and $\theta$, which we do not show explicitly because of their cumbersome form, but that have been used in the paper calculations and especially in the numerics.

{Although we have already reached our goal of calculating the moduli $\tilde E$ and $\tilde B$ (via the NP formalism), which are the quantities we need to calculate the pair creation rate (see Eq. \ref{eq:rate2} in Section \ref{sec:2}), we now present for completeness} an alternative derivation by applying a Lorentz boost to the ZAMO, which moves the frame to one where the fields are parallel to each other. We denote the boost as ${\bf \Lambda}(v)$, with $\vec{v} = v e_{\hat 3}$, defined implicitly by (see, e.g., \citealp{1975ctf..book.....L})
\begin{equation}\label{eq:boost}
    \frac{v}{1+v^2} \equiv \sigma^{-1} = \frac{|{\bf \hat E}\times{\bf \hat B}|}{\hat{E}^2 + \hat{B}^2} = \frac{E_{\hat{1}}B_{\hat{2}}-E_{\hat{2}}B_{\hat{1}}}{\hat{E}^2 + \hat{B}^2}.
\end{equation}
We can solve Eq. (\ref{eq:boost}) for the boost speed as
\begin{equation}\label{eq:v}
    v = \frac{\sigma - \sqrt{\sigma^2-4}}{2},
\end{equation}
where the sign is chosen to have the correct physical underluminal solution. The electric and magnetic fields in the new frame are ${\bf \tilde E} = E^{(i)} e_{(i)} = E_{(i)} e^{(i)}$ and ${\bf \tilde B} = B^{(i)} e_{(i)} = B_{(i)} e^{(i)}$, with $E_{(i)} = \Lambda_{(i)}^{\hphantom{(i)}{\hat{j}}}E_{\hat{j}}$ and $B_{(i)} = \Lambda_{(i)}^{\hphantom{(i)}{\hat{j}}}B_{\hat{j}}$, being $\Lambda_{(i)}^{\hphantom{(i)}{\hat{j}}}$ is the Lorentz transformation. It leads to
\begin{subequations}\label{eq:EBparcomp}
    \begin{align}
    E_{(1)} &= \gamma (E_{\hat{1}} - v B_{\hat{2}}),\quad E_{(2)} = \gamma (E_{\hat{2}} + v B_{\hat{1}}),\\
    B_{(1)} &= \gamma (B_{\hat{1}} + v E_{\hat{2}}),\quad B_{(2)} = \gamma (B_{\hat{2}} - v E_{\hat{1}}),
\end{align}
\end{subequations}
where $\gamma = 1/\sqrt{1-v^2}$. With the above, the intensity of the electric and magnetic field in the new frame are
\begin{subequations}\label{eq:EBparmodule}
\begin{align}
    \tilde{E} &= \sqrt{E_{(1)}^2 +  E_{(2)}^2} = \gamma \sqrt{\hat{E}^2 + v^2 \hat{B}^2 -2 v |{\bf \hat E}\times{\bf \hat B}|},\label{eq:Eparmodule}\\
    \tilde{B} &= \sqrt{B_{(1)}^2 +  B_{(2)}^2} = \gamma \sqrt{\hat{B}^2 + v^2 \hat{E}^2 -2 v |{\bf \hat E}\times{\bf \hat B}|},\label{eq:Bparmodule}
\end{align}
\end{subequations}
which can be checked, lead to the same expressions obtained above with the NP formalism.

By inserting the boost speed given by Eq. (\ref{eq:v}) into Eq. (\ref{eq:EBparmodule}), we have analytic (though cumbersome) expressions of $\tilde E$ and $\tilde B$ as a function of the ZAMO field components $E_{\hat i}$ and $B_{\hat i}$. The latter are given in Eq. (\ref{eq:EMfieldzamo}).

We can explicitly see the difference relative to the LNRF fields in the weak-field regime. By performing a $1/r$ power expansion of Eq. (\ref{eq:EBparmodule}), defining the dimensionless coordinate $\bar r \equiv r/M$ and $\xi \equiv a/M$, we obtain
\begin{subequations}\label{eq:EBparexpansion}
    \begin{align}
     \tilde E &\simeq \frac{\xi B_0}{\bar r^2} |\cos\theta (3 \cos^2\theta-1)|= \hat E |\cos\theta|,\\
     \tilde B &\simeq B_0 \left(1-\frac{\sin^2\theta}{\bar r} \right) = \hat B.
\end{align}
\end{subequations}
Equation (\ref{eq:EBparexpansion}) shows that, at first order, $\tilde B = \hat B$. Further, along the polar axis ($\theta = 0$), $\tilde E = \hat E$ and $\tilde B = \hat B$. The latter is true in general since the electric and magnetic fields in the LNRF are parallel along $\theta = 0$, as can be checked from Eq. (\ref{eq:EB}), so $v = 0$ on the polar axis.

It is worth mentioning that the electric and magnetic fields in the LNRF frame and the parallel fields are, quantitatively speaking, nearly indistinguishable. The reason is that the Lorentz boost relating the two frames given by Eq. (\ref{eq:v}) is weakly relativistic, i.e., it differs from unity by less than $0.1\%$ for relevant astrophysical parameters.

%\bibliographystyle{aa}
%\bibliography{references}

\end{document}